\documentclass{rsproca}

\usepackage{amsmath,amsfonts,amssymb,amsthm,amsbsy,,dsfont,mathtools}

\newcommand{\ket}[1]{\mathop{\left|#1\right>}\nolimits}            
\newcommand{\kbd}{\mathop{\left|\bullet\right>}\nolimits}            

\newcommand{\nn}{\nonumber}

\def\ddg{\ddagger}

\newcommand{\brk}[2]{\langle #1 | #2 \rangle}

\newcommand{\bu}{\bullet}

\def\a{\alpha}
\def\b{\beta}
\def\g{\gamma}
\def\Ga{\Gamma}
\def\d{\delta}

\def\vp{\varphi}

\def\G{\mathcal{G}}

\DeclareMathOperator{\OSp}{OSp}
\DeclareMathOperator{\UOSp}{UOSp}
\DeclareMathOperator{\SU}{SU}
\DeclareMathOperator{\Un}{U}

\begin{document}

\title{Tsirelson's bound and supersymmetric entangled states}

\author{
L. Borsten$^{1}$, K. Br\'adler$^{2}$ and M. J. Duff$^{1}$}

\address{$^{1}$Theoretical Physics,
    Blackett Laboratory,
    Imperial College London,
    London SW7 2AZ, United Kingdom\\
$^{2}$School of Computer Science,
    McGill University,
    Montreal, Quebec, H3A 2A7, Canada}

\subject{11.30.Pb, 03.65.Ud, 03.67.Mn}


\corres{L. Borsten\\
\email{leron.borsten@imperial.ac.uk}}

\begin{abstract}
A superqubit,  belonging to a $(2|1)$-dimensional super-Hilbert space,  constitutes the minimal supersymmetric extension of the conventional qubit. In order to see whether superqubits are more nonlocal than ordinary qubits, we construct a class of two-superqubit entangled states as a nonlocal resource in the CHSH game. Since super Hilbert space amplitudes are Grassmann numbers, the result depends on how we extract real probabilities and we examine three choices of map: (1) DeWitt (2) Trigonometric (3) Modified Rogers. In cases (1) and (2) the winning probability reaches the Tsirelson bound $p_{win}=\cos^2{\pi/8}\simeq0.8536$  of standard quantum mechanics. Case (3) crosses Tsirelson's bound with $p_{win}\simeq0.9265$. Although  all states used in the game involve probabilities lying between 0 and 1, case (3) permits other changes of basis inducing negative transition probabilities.
\end{abstract}


\begin{fmtext}

\section{Introduction}

 In providing nonlocal resources, quantum mechanics fundamentally distinguishes itself from the  classical world. Perhaps the best known example of a nonlocal resource is the EPR-pair, introduced by Einstein, Podolsky and Rosen \cite{Einstein:1935rr}. Bell famously showed  that this quantum state can be used to violate  an inequality required to be satisfied by any local hidden variable model \cite{Bell:1964kc}. It was later realised  by Tsirelson that not only does the EPR-pair violate the Bell inequality, it violates it maximally  \cite{cirel1980quantum}: there is no quantum system that can do better.  In particular, Tsirelson used a variant of the Bell inequality introduced by Clauser, Horne, Shimony and Holt (CHSH) \cite{Clauser:1969}. The CHSH inequality can be nicely rephrased  as  a limit on the probability of  winning a \emph{nonlocal game} \cite{1313847}, which we describe below. In these terms the CHSH inequality is given by $p_{win}\leq3/4$, where $p_{win}$ denotes the probability of winning the  game. 
  \end{fmtext}

\maketitle

 On the other hand Tsirelson's bound corresponds to  $p_{win}\leq\cos^2{\pi/8}\simeq0.8536$.  Why does quantum mechanics stop at $p_{win}=\cos^2{\pi/8}$? In an effort to address this very question Popescu and Rohrlich \cite{popescu1994quantum} asked whether one could envision a theory which beats $p_{win}=\cos^2{\pi/8}$, i.e. is more nonlocal than quantum mechanics, without violating other established principles of physics, in particular, the no-signalling condition imposed by special relativity.  They answered this question in the positive: there are hypothetical resources \cite{popescu1994quantum}, known as {\it nonlocal-boxes} or \emph{PR-boxes}, that have $p_{win}=1$ without violating the no-signalling condition. In  \cite{PhysRevA.75.032304} it was shown that PR-boxes are  realised in a generalised probabilistic theory, often referred to as box-world, which makes use of generalised bits.  However, they  are purely mathematical constructs with no links to existing physical theories. 

\section{Superqubits and Tsirelson's bound}

In this paper we ask whether {\it superqubits} can cross Tsirelson's bound. Superqubits are a supersymmetric extension of qubits, introduced in \cite{Borsten:2009ae}, where the  even (commuting) computational basis vectors $\ket{0}$ and $\ket{1}$ are augmented by an odd (anticommuting) basis vector $\ket{\bu}$. They transform as a triplet under  the orthosymplectic group $\UOSp(1|2)$ (the ``compact real form'' of $\OSp(1|2)$), which is the minimal supersymmetric extension of the $\SU(2)$ group of local unitaries. The $\mathfrak{osp}(1|2)$ algebra has been extensively studied in the past \cite{landi1987extensions,chaichian:3381,Grosse:1997fk,bartocci:45} in the context of Lie superalgebras \cite{Rittenberg:1978bu,kac1977lie,scheunert1979theory}. In general, the main physical motivation for studying supersymmetry comes from high energy physics where it is a leading candidate for physics beyond the Standard Model. However, as stressed in \cite{Borsten:2009ae}, the superqubit does  not belong to this branch of ideas. Firstly, it is not a representation of the super-Poincar\'e  group and, secondly, it is a non-trivial extension of conventional quantum mechanics based on a super Hilbert space \cite{DeWitt:1984, rogers:1352, rudolph-2000-214}.  It is worth remarking, however, that supersymmetry also appears in several proposals from condensed matter physics. In particular, the orthosymplectic algebra $\mathfrak{osp}(1|2)$ plays a key role in \cite{Hasebe200594,Hasebe2011777} and superqubit basis states make an appearance in the context of  supersymmetric valence-bond solid states in \cite{PhysRevB.84.104426, hasebe2013topological}.

We begin by recalling the usual CHSH game \cite{Clauser:1969}. It is a so-called nonlocal game \cite{1313847,RevModPhys.82.665} with three players: a referee who competes with Alice and Bob. The referee chooses with uniform probability $1/4$ two bits $i\in\{0,1\}$ and $j\in\{0,1\}$ and sends $i$  to Alice and $j$ to Bob such they are not aware of one another's bit value. Alice and Bob each send one bit, denoted $a$ and $b$ respectively, back to the referee. The conditions for Alice and Bob to win the game are captured in the following table:
\begin{center}
    \begin{tabular}{c|c}
      $ij$ & $a\oplus b$ \\
      \hline
      00 & 0 \\
      01 & 0 \\
      10 & 0 \\
      11 & 1 \\
\end{tabular}
\end{center}
Here, $\oplus$ denotes addition mod 2. Alice and Bob cannot communicate during the game but they may establish their strategy beforehand. Tsirelson's bound is achieved if they share a maximally entangled state such as $\Psi=(\ket{00}+\ket{11})/\sqrt{2}$ accompanied by an agreed measurement strategy. The PR-box   always   wins -  it is simply a black-box that when given $i$ by Alice and $j$ by Bob  outputs $a, b$ such that the table is satisfied \cite{popescu1994quantum}. The important point is that it still respects the no-signalling principle. 

A normalized superqubit $\psi$
\cite{Borsten:2009ae,Castellani:2010yz} may be regarded as an element of the projective space $S^{2|2}=\UOSp(1|2)/\Un(0|1)$ known as the {\it supersphere}~\cite{landi1987extensions,chaichian:3381,Grosse:1997fk,bartocci:45,Hasebe200594}, where $\Un(0|1)$ is the even Grassmann generalization of the $\Un(1)$ Lie group. The definition implies that the supersphere is nothing other than a super version of the Bloch sphere $S^2=\SU(2)/\Un(1)$.
An arbitrary superqubit can be generated by a general group action
\begin{align}\label{eq:UOSPgroupmanifold}
    &Z(\eta,\a,\b)
     =S(\eta)U(\a,\b)
     =\begin{pmatrix}
       1+{1\over4}\eta\eta^\# & -{\eta\over2} & {\eta^\#\over2} \\
       -{\eta^\#\over2} & 1-{1\over8}\eta\eta^\# & 0 \\
       -{\eta\over2} & 0 & 1-{1\over8}\eta\eta^\#
      \end{pmatrix}
      \begin{pmatrix}
        1 & 0 & 0 \\
        0 & \a & -\b^\# \\
        0 & \b & \a^\#
      \end{pmatrix},
\end{align}
where $\a,\b$ are even Grassmann supernumbers satisfying $\a\a^\#+\b\b^\#=1$ and $\eta$ is an unconstrained odd supernumber. Supernumbers are  elements of a complex Grassmann algebra which is a complex vector space equipped with an alternating product. A Grassmann algebra is isomorphic to an exterior algebra. The hash operator denotes the graded involution which is a generalized complex conjugation. For  even supernumbers $\alpha^{\#\#}=\alpha$, for  odd supernumbers $\eta^{\#\#}=-\eta$ and for arbitrary supernumbers $(ab)^\#=a^\#b^\#$.

Using Eq.~(\ref{eq:UOSPgroupmanifold}) a general superqubit state is given by
\begin{equation}\label{eq:superqubit}
       \ket{\psi}=\left(1-{1\over8}\eta\eta^\#\right)(\a\ket{0}+\b\ket{1})+{1\over2}(-\a\eta+\b\eta^\#)\kbd.
\end{equation}
The involution is used to define the graded adjoint $X^\ddagger\overset{\rm df}{=}(X^{ST})^\#$ where $ST$ stands for the supertranspose~\cite{Varadarajan:2004,kac1977lie}.
The graded adjoint acts as ordinary adjoint for even linear operators and vectors but $(X^\ddg)^\ddg=-X$ holds for odd operators and vectors~\cite{Rittenberg:1978bu,scheunert1979theory}. We further define the Grassmann-valued transition probability function between two superqubits $\vp$ and $\psi$ as
\begin{equation}\label{eq:born}
    p_\G(\vp,\psi)=\brk{\vp}{\psi}\big(\brk{\vp}{\psi}\big)^\#.
\end{equation}
The rationale behind this definition is clear: for ordinary (non-Grassmann) states we recover the usual Born rule. Furthermore, for any superqubit $\psi$ we find that $\sum_{m=1}^3p_\G(m,\psi)=1$ where $m=\{0,1,\bu\}$. Hence all superqubits are normalized to one. Adopting the rule of one Grassmann generator $\theta$ per superqubit the single superqubit outcome probabilities in the computational basis are given by
\begin{subequations}\label{eq:prob}
\begin{align}
    p_{\G0}&=\alpha\bar\alpha(1-r^2\theta\theta^\#)\\
    p_{\G1}&=\b\bar\b(1-r^2\theta\theta^\#)\\
    p_{\G\bu}&=r^2\theta\theta^\#,
\end{align}
\end{subequations}
where we set $\eta=2r\theta$ in Eq.~(\ref{eq:superqubit}) ($r\in\mathds{R}$). The parameters $\a,\b$ are in general even Grassmann numbers but we regard them as complex for simplicity. 

We now require a final map to convert the Grassmann-valued probabilities $p_{\G m}$ into a real-valued probabilities, $p_m$, which respects the $\UOSp(1|2)$ symmetry. 
Note, the supersymmetric extension of the Born rule given in \eqref{eq:born} is not necessarily the unique consistent choice. However, given its evident naturalness and consistent reduction to the standard Born rule, we will not consider other possibilities  here \footnote{There is a considerable literature on the possible consistent extensions/generalisations of quantum mechanics and their  implications for information processing.   See, for example, \cite{PhysRevA.75.032304, PhysRevLett.99.240501}. It should be noted, however, that we are not considering a generic point in this space as our construction is highly constrained by the symmetries of the superqubit.}. On the other hand there is clearly more freedom in choosing the map taking the Grassmann-valued probabilities into the reals, which has no counter part in  standard quantum mechanics.  Here, we  consider  three possible alternatives. 

The first map, introduced by DeWitt \cite{DeWitt:1984}, is to simply ignore the Grassmann part of the probability. Every supernumber may be uniquely split into a pure  complex number and a sum over the Grassmann algebra generators with complex coefficients, often referred to as the \emph{body} and \emph{soul}, respectively,
\begin{equation}
z=z_{body} + z_{soul}, 
\end{equation}
where $z_{body}\in\mathds{C}$. The DeWitt map simply sends $p_\G$ into $ p_{\G body}$.  It boils down to the ordinary quantum mechanics of a qubit: $p_{0}=\alpha\bar\alpha,p_{1}=\b\bar\b$ and $p_{\bu}=0$ where the bar denotes complex conjugation. 

The second possibility is the trigonometric map: $r^2\theta\theta^\#\mapsto\cos^2{r}$. Hence $p_{0}=\a\bar\a\sin^2{r},p_{1}=\b\bar\b\sin^2{r}$ and $p_{\bu}=\cos^2{r}$ and $p_{0}+p_{1}+p_{\bu}=1$. Evidently,   these maps share the virtue that the individual probabilities lie between zero and one.  

The third map we will explore is defined by the rule $r^2\theta\theta^\#\mapsto r^2$, where we fix the orientation of the Grassmann generators to be $\theta\theta^\#$. Hence $p_{0}=\alpha\bar\alpha(1-r^2),p_{1}=\b\bar\b(1-r^2)$ and $p_{\bu}=r^2$. We refer to this map as the modified Rogers map since, in spite of their similarity, the Rogers construction ~\cite{rogers:1352,rudolph-2000-214} does not respect the $\UOSp(1|2)$ symmetry. For a $2^{2n}$-dimensional Grassmann algebra generated by $\{\theta,\theta^\#\}_{i=1}^n$ the definition of the modified Rogers map can be generalized as follows
\begin{equation}\label{eq:Grassnorm2}
\big|\tau\big|_{R}\overset{\rm df}{=}\int  \prod^{n}_{i}e^{\theta_i\theta^\#_{i}}\tau {\rm d}^{2n}\theta,
\end{equation}
where $\tau$ is an arbitrary supernumber, ${\rm d}^{2n}\theta=\prod_{i=1}^n{\rm d}\theta_i{\rm d}\theta_i^\#$ and the orientation is fixed by $\prod_{i=1}^n\theta_i\theta_i^\#$. The map is linear and respects the $\UOSp(1|2)$ symmetry. However, although the probabilities sum to one, they are not always guaranteed individually to lie between zero and one.

We are ready to play the superqubit CHSH game in each of the three cases, being careful in case (3) to avoid measurement outcomes with negative probabilities. A superqubit is formally a three-level system and so we have to adapt the strategy employed by the players  to this situation. Alice announces $a=0$ ($a=1$) to the referee if her local measurement projects to $\ket{0}$ ($\,\ket{1}$ or $\ket{\bu}$). The same holds for the value of $b$ transmitted from Bob's laboratory.
 Therefore the winning Grassmann-valued probability reads
\begin{equation}\label{eq:grassWINprob}
    p_{\G}={1\over4}
    \Bigg[ \sum_{ij\in\{00,01,10\}}
    \Big( p_{\G{00}}^{(ij)}+ p_{\G11}^{(ij)}+ p_{\G1\bu}^{(ij)}+ p_{\G\bu1}^{(ij)}+ p_{\G\bu\bu}^{(ij)}\Big)+p_{\G01}^{(11)}+ p_{\G10}^{(11)}+ p_{\G0\bu}^{(11)}+ p_{\G\bu0}^{(11)}\Bigg],
\end{equation}
where subscripts $A$ and $B$ refer to Alice and Bob and 
\begin{equation}\label{eq:probfunction}
     p^{(ij)}_{\G mn}=\brk{m_{A}n_{B}}{\Ga_{iA,jB}}\big(\brk{m_{A}n_{B}}{\Ga_{iA,jB}}\big)^\#
\end{equation}
is the Grassmann-valued probability function introduced in Eq.~(\ref{eq:born}). $\Ga_{iA,jB}=(Z_{iA}\otimes Z_{jB}){\Ga}_{AB}$ is a locally superunitary rotated superqubit bipartite entangled state $\Ga_{AB}$ where
\begin{equation}\label{eq:localsuperrotation}
    Z_{iA}\otimes Z_{jB}=S(2r_i\theta_A)U(\a_i,\b_i)\otimes S(2s_j\theta_B)U(\g_j,\d_j).
\end{equation}
Here, the Grassmann odd elements $\theta_A,\theta_B$ and their conjugates are the generators of the Grassmann algebra of order $2n=4$ and $r_i,s_j\in\mathds{R}$. The shared entangled state we use reads
\begin{equation}\label{eq:Gamma}
    \Ga_{AB}=\Big(1+{1\over2}n+{3\over8}n^2\Big)\Big({1\over\sqrt{2}}(\ket{00}+\ket{11})
    +{p\over\sqrt{2}}\theta_A\ket{\bu1}+{q\over\sqrt{2}}\theta_B\ket{1\bu}\Big)
\end{equation}
where $p,q\in\mathds{R}$ and $n=-{p^2\over2}\theta_A\theta^{\#}_A-{q^2\over2}\theta_B\theta^{\#}_B$. The form of $n$ is  fixed by the normalisation condition $\langle \Ga_{AB}|\Ga_{AB}\rangle=1$. Note that this state reduces to the Bell state in the absence of the Grassmann numbers ($p=q=0$). 

We define
\begin{equation}\label{eq:maxprocedure}
         p_{win}=\max_{\substack{p,q,r_i,s_j \\ \a_i,\b_i,\g_j,\d_j}}{p_{win}(\Ga_{AB})}\quad
         \text{s.t.} \quad
         0\leq p^{(ij)}_{mn}\leq1,\ \ \ \forall i,j,m,n
\end{equation}
where $p_{win}(\Ga_{AB})$ is obtained from Eq.~(\ref{eq:grassWINprob}) on applying one of the three alternative maps from the Grassmanns to the reals. For both the DeWitt and trigonometric maps Tsirelson's bound is met but not exceeded. Note, this does not exclude the possibility that the bound may be crossed using a different state or  strategy.  However, in the case of the modified Rogers map Tsirelson's bound is explicitly shown to be crossed even when restricting to  measurements that cannot result in  negative probabilities for the given state.  The optimization procedure has to be performed numerically~\cite{1393890}. The winning parameters are
    \begin{align}\label{eq:winningparameters_rs}
        &r_0\simeq0.7476,\ s_0\simeq 0.6329,\ r_1\simeq 0,\ s_1\simeq 0.6329\\\nn
        &\a_0\simeq-\pi/2,\a_1\simeq\pi/4,\b_0\simeq\pi/4,\b_1\simeq3\pi/4\\
        &p\simeq0.7476,\ q\simeq-1.0949\nn
    \end{align}
This corresponds to $p_{win}\simeq0.9265$. Note that this is a mere lower bound since the optimization problem is nonconvex. We have not  determined the upper bound on $p_{win}$. We appreciate that this exceeds  the triviality of communication complexity bound $p_{win}\leq (3+\sqrt{6})/6\simeq 0.908$ \cite{2005quant.ph..1159V, PhysRevLett.96.250401}. However, a case can be made for imposing $|r_i|\leq1/2, |s_j|\leq1/2$, which is computationally equivalent to `compactifying' $S\in\UOSp(1|2)$ appearing in  (\ref{eq:localsuperrotation}).  In this case we obtain a reduced $p_{win}=0.8647$, respecting the triviality of communication complexity bound. For a more detailed discussion of this approach the reader is referred to \cite{Bradler:2012ii}. In \cite{pawlowski2009information} the \emph{principle of information causality} was proposed as basic property of Nature. It generalises the no-signalling principle and distinguishes quantum theory from no-signalling theories with stronger than quantum correlations.   Even in the latter case  superqubits, using the modified Rogers map,   violate the principle of information causality, since the   Tsirelson bound for the CHSH inequality is exceeded. 

\section{Conclusions}

In this work we have proposed  a  model based on supersymmetry which provides bipartite entangled  states more nonlocal than those allowed by quantum mechanics.

Violating Tsirelson's bound was always destined to involve paying a price and it remains to be seen whether the existence of negative transition probabilities is too high price to pay (even though we did not invoke them in the CHSH game). In fact, one
might ask whether it is this feature alone, with or without supersymmetry, that is responsible for exceeding the bound. Indeed, it has been shown that the space of no-signalling models is equivalent to the space of local hidden-variable models with extended probabilities that marginalize to yield standard non-negative
probabilities \cite{abramsky2011sheaf}. However, extended probabilities alone do not seem sufficient. Evidence for this is provided by the example of a `qutrit' with non-compact local unitary group $\SU(2,1)$, rather than $\SU(3)$, which, like the superqubit, has an $\SU(2)$ subgroup. The probabilities  sum to one but, owing to the indefinite metric, are again not always guaranteed individually to lie between zero and one. Unlike the superqubit, however, there are no rotations, aside from the trivial $\SU(2)$ subgroup, that do not lead to negative transition probabilities. Excluding such rotations, therefore, means that our `qutrits' cannot exceed the bound. This  suggests that the violation of Tsirelson's bound  in the case of two superqubits should not  be attributed to extended  probabilities alone.

\section*{Acknowledgment}

We are grateful to the The Royal Society for an International Exchanges travel grant and to Samson Abramsky, Duminda Dahanayake, P\'eter L\'evay, Markus M\"uller and William Rubens for illuminating discussions and help with Mathematica.
KB acknowledges support from the Office of Naval Research (grant No. N000140811249). KB and MJD are grateful for hospitality at the Aspen Center for Physics and the Centre for Quantum Technology at the University of KwaZulu-Natal during NITheP Workshop (Relativistic Quantum Information) where  part of this work was done. The work of LB is supported by an Imperial College Junior Research Fellowship and in part by ERC
advanced grant no. 226455, Supersymmetry, Quantum Gravity and Gauge Fields (SUPERFIELDS). The work of MJD is supported by the STFC under rolling grant ST/G000743/1.

\providecommand{\href}[2]{#2}\begingroup\raggedright\endgroup

\end{document}